\documentclass{elsart}

\usepackage{amssymb}
\usepackage{graphicx}

\begin{document}

\newcommand{\cal}{\mathcal}

\newcommand{\eg}{{\it e.g. \/}}

\newcommand{\ie}{{\it i.e. \/}}

\newcommand{\cf}{{\it cf}\/}
\newcommand{\dg}{$^\circ$}
\newcommand{\saut}{\vspace*{5 mm}\par}

\newcommand{\RR}{\hbox{\bf I\hspace*{-1mm}R}}

\newcommand{\ds}{\displaystyle}

\begin{frontmatter}



\title{A reduced-order strategy for 4D-Var data assimilation}


\author{C. Robert\corauthref{cor1}},
\author{S. Durbiano},
\author{E. Blayo},
\author{J. Verron},
\author{J. Blum},
\author{F.-X. Le Dimet}
\corauth[cor1]{Corresponding author. Email address : Celine.Robert@imag.fr}
\address{IDOPT Project, LMC-IMAG and INRIA Rh\^{o}ne-Alpes, BP 53X, 38041 Grenoble cedex, France}
\begin{abstract}
This paper presents a reduced-order approach for four-dimensional variational data assimilation, based on a prior EOF analysis of a model trajectory. This method implies two main advantages: a natural model-based definition of a multivariate background error covariance matrix $\textbf{B}_r$, and an important decrease of the computational burden of the method, due to the drastic reduction of the dimension of the control space.
An illustration of the feasibility and the effectiveness of
this method is given in the academic framework of twin experiments for a
model of the equatorial Pacific ocean. It is shown that the multivariate aspect of $\textbf{B}_r$ brings additional information which substantially improves
the identification procedure. Moreover the computational cost can be decreased by one order of magnitude with regard to the full-space 4D-Var method.
\end{abstract}
\end{frontmatter}
\section{Introduction}
The aim of this paper is to investigate a reduced-order approach for four-dimensional variational data assimilation (4D-Var), with an illustration in the context of ocean modelling, which is our main field of interest.
4D-Var is now in use in
numerical weather prediction centers (e.g. Rabier {\it et al.} 2000) and should be a potential
candidate for operational oceanography in prospect of seasonal climate
prediction and possibly of high resolution global ocean mesoscale prediction.
However, ocean scales make the problem even more difficult
and computationally heavy to handle than for the atmosphere.
Several applications were conducted these last years for various oceanic
studies, including for example : basin-scale ocean circulation, either with
quasigeostrophic (Moore 1991; Schr\"oter {\it et al.} 1993; Luong {\it
et al.} 1998) or with primitive equation models (Greiner {\it et al.}
1998; Wenzel and Schr\"oter 1999; Greiner and Arnault 2000; Weaver {\it et al.}
2002); coastal modelling (Leredde {\it et al.} 1998; Devenon {\it et
al.} 2001); or biogeochemical modelling (Lawson {\it et al.} 1995;
Spitz {\it et al.} 1998; Lellouche {\it et al.} 2000, Faugeras {\it et
al.}, 2003).\par
However, although considerable work and improvements have been performed,
a number of difficulties remain, common to most applications (and also to other data assimilation methods). The first
problem is the fact that ocean models are non-linear, while 4D-Var theory is
established in a linear context. More precisely, variational approach can
adapt in principle to non-linear models, but the cost function is no
longer quadratic with regard to the initial condition (which is the
usual control parameter) which can lead to
important difficulties in the minimization process and the occurence of
multiple minima. Several strategies have been proposed to overcome these
 problems: Luong {\it et al.} (1998) and Blum {\it et
al.} (1998) perform successive minimizations over increasing time periods; 
Courtier {\it et al.} (1994), with the so-called incremental approach, generate a
succession of quadratic problems, which solutions should converge (but with no
general theoretical proof) towards the solution of the initial minimization
problem. 
A second major difficulty with variational problem
implementation lies in our poor knowledge of the
background error, whose covariance matrix plays an important role in the cost
function and in the minimization process. In the absence of statistical information, these covariances are
often approximated empirically by analytical (e.g. Gaussian)
functions. For instance, the covariances, used in the ``standard''
4D-Var experiment $E_{FULL}$ described in section \ref{Sect:NumExp} are 3D but univariate.
Moreover, as discussed in (Lermusiaux, 1999), errors evolve with the
dynamics of the system and thus the error space should evolve in the
same way. In realistic systems, it proves to be difficult to catch correctly this evolution.
The third major problem in the use of 4D-Var in realistic oceanic applications
is probably the dimension of the control space.  In fact, this
dimension is generally equal to the size of the model state
variable (composed, in our case, by the two horizontal components of the
velocity, temperature and salinity), which is
typically of the order of $10^6$-$10^8$. This makes of course the
minimization difficult and expensive (typically tens to hundreds times the
cost of an integration of the model), even with the best current
preconditioners.\par 
This last difficulty can be addressed by reducing the dimension of the
minimization space. This is for example the idea of the incremental approach
(Courtier {\it et al.} 1994), in which an important part of the
successive quadratic minimization problems previously mentioned can be
solved using a coarse resolution (e.g. Veers\'e and Th\'epaut 1998).
The dimension of the minimization problem can then be decreased by one
or two orders of magnitude. However, even with such an approach, the
dimension of the control space remains quite large in realistic
applications. Another way to reduce the dimension of the control space
 is the representer method (Bennett, 92), performing the minimization
in the observation space. The number of parameters to estimate is
equal to the number of observation locations.
Concerning sequential data assimilation, reduced-order methods were
developed to allow the specification of error covariances matrix even
for realistic applications. This is the case for example of the
Singular Extended Evolutive Kalman (SEEK)
filter (Pham {\it et al.} 1998; Brasseur {\it et al.}
1999). \par
In this paper, we propose an alternative way for drastically decreasing the dimension of the control space, and hence the cost of the minimization process. Moreover this method provides a natural choice for a multivariate background error covariance matrix, which helps improving the quality of the final solution. The method is based on a decomposition of the control variable on a well-chosen family of a few relevant vectors, and has already been successfully applied in the simple case of a quasigeostrophic box model (Blayo {\it et al.} 1998). The aim of the present paper is to further develop this approach and to validate it in a more realistic case, namely a primitive equation model of the equatorial Pacific ocean. The method is described in section 2. Then the model, the assimilation scheme and the numerical experiments are presented in section 3, and their results are discussed. Finally some conclusions are drawn in section 4.
\section{The reduced-space approach}
Let a model simply written as
\begin{equation}
\frac{\partial {\bf x}}{\partial t} = M({\bf x}) 
\end{equation}
with the state vector ${\bf x}$ in $\Omega
\times [t_0,t_N]$, $\Omega$ being the physical domain. Suppose that we have some
observations ${\bf y}^{\hbox{\scriptsize o}}$ distributed over $\Omega \times
[t_0,t_N]$, with an observation operator $H$ mapping ${\bf x}$ onto ${\bf y}$. The classical 4D-Var approach consists in minimizing a cost function 
\begin{equation}
\begin{array}{ll}
J({\bf u})&= J_o({\bf u}) + J_b({\bf u})\\
&= \frac{1}{2} \,\displaystyle \sum_{i=0}^N \left( H({\bf x}_i) - {\bf y}_i^{\hbox{\scriptsize o}} \right)^T {\bf R}_i^{-1} \left( H({\bf x}_i) - {\bf y}_i^{\hbox{\scriptsize o}} \right) + \frac{1}{2} \, ( {\bf u} - {\bf u}^b )^T {\bf B}_u^{-1} ( {\bf u} - {\bf u}^b )
\end{array}
\label{eq:J}
\end{equation}
using the notations of Ide {\it et al.} (1997). ${\bf u}^b$ is a background value for the control vector ${\bf u}$, and ${\bf B}_u$ is its associated error covariance matrix. In most applications, the control variable ${\bf u}$ is the state variable at
the initial time : ${\bf u}={\bf x}(t_0)$, and the background state
${\bf u}^b = {\bf x}^b$ is typically a forecast from a previous
analysis given by the data assimilation system. In this case, once the model is discretized,
the size of ${\bf u}$ (i.e. the dimension of the control space ${\cal
U}$) is equal to the size of ${\bf x}$, denoted by $n$. $\ \textbf{x}_i$ stands for the state variable at time $t_i$. In equation (\ref{eq:J}), $\textbf{x}_i$ is propagated by $M$, the fully non-linear model.\par
In the incremental formulation which is used here, the cost function ${\bf{J}}$ is written as
a function of  $\delta {\bf x}_0 = {\bf x}_0 - {\bf x}^b$ and the $J_o$ term is calculated using the linearized model \textbf{M}:

\begin{equation}
\begin{array}{ll}
J(\delta \textbf{x}) & =\frac{1}{2}(\delta \textbf{x})^t \textbf{B}^{-1}\delta
\textbf{x} \\
  &  + \frac{1}{2} \displaystyle \sum_{i=1}^{N} ({\bf{H_{i}M_{t_i,t_0}}}\delta
\textbf{x}_0 -\bf{d}_i)^{t} R_{i}^{-1} ({\bf{H_{i}M_{t_i,t_0}}}\delta
\textbf{x}_0 -\bf{d}_i) \\
\end{array}
\label{eq:Jinc}
\end{equation}

where $\bf{d}_i$ stands for the innovation vector:  $\textbf{d}_i =
\textbf{y}_i - H(\textbf{x}_b(t_i))$ and $M_{t_i,t_0}$ is the temporal
evolution performed by the model $M$ between the instants $t_0$ and $t_i$.\par
The basic idea then, for constructing a reduced-order approach, consists in
defining a convenient mapping ${\cal M}$ from ${\cal W}\equiv\RR^r$ into
$\,{\cal U}\equiv\RR^n$, with $r\ll n$, and in replacing the control variable ${\bf u}$ by the new control variable ${\bf w}$ with ${\bf u}={\cal M}({\bf w})$. Since we want to preserve a good solution while having only a rather small number $r$ of degrees of freedom on the choice of ${\bf w}$, the subspace ${\cal M}({\cal W})$ of ${\cal U}$ must be chosen in order to contain only the ``most pertinent'' admissible values for ${\bf u}$. More precisely, in the case of the control of the initial condition ${\bf u}={\bf x}(t_0)$, we decide to define the mapping ${\cal M}$ by an affine relationship of the form :
\begin{equation}
{\bf x}(t_0)={\cal M}({\bf w})=\hat{{\bf x}} + \sum_{i=1}^r w_i {\bf L}_i \qquad \hbox{with } {\bf w}=(w_1,\ldots,w_r) \in
{\cal W}\equiv\RR^r 
\label{eq:x0}
\end{equation}
In order to let ${\bf w}$ span a wide range of physically possible
states, $\hat{{\bf x}}$ represents an estimate of the state of the
system, and  ${\bf L}_1,\ldots, {\bf L}_r$ are vectors containing the
main directions of variability of the system (the $w_i$ are scalars).
Such a definition relies on the fact that most of the variability of
an oceanic system can be described by a low dimensional space.
Even if it is only rigorously proved for very simplified models (Lions
\textit{et al.}, 1992), it is often
expected that, away from the equator, ocean circulation can be seen as a dynamical
system having a strange attractor. This means that the system trajectories are
attracted towards a (low dimension) manifold. In the vicinity of this
attractor, orthogonal perturbations will be naturally damped, while tangent
perturbations will not (they can even be greatly amplified, due to the chaotic
character of the system). To retrieve a system trajectory over of period of
time $[t_0,t_N]$, it seems thus necessary to propose an initial condition ${\bf x}(t_0)$
containing such variability modes tangent to the attractor, but not
necessarily variability modes orthogonal to it. Thus, in definition
(\ref{eq:x0}), $\hat{{\bf x}}$ should ideally be located on the attractor,
and ${\bf L}_1,\ldots, {\bf L}_r$ should correspond to the main directions of
variability tangent to it. In the tropical ocean, the rationale is different, and even simpler since the tropical ocean dynamics is mostly linear, and can be represented by a rather limited number of linear, and possibly non-linear, modes (e.g. De Witte {\it et al.} 1998).\par 
In practice, we will choose $\hat{{\bf x}}={\bf x}^b$, i.e. the background state that would be used in the corresponding classical 4D-Var approach. 
With this choice, the increment $\delta {\bf x}= {\bf x}(t_0)-{\bf x}^b$ is equal to $\ds{\delta {\bf x}= \sum_{i=1}^r w_i {\bf L}_i = {\bf L w}}$. In this reduced-space approach, we define a new expression for the background term $J_b$ of the cost function $J$ :
\begin{equation}
J_b({\bf w})=\frac{1}{2} \, {\bf w}^T {\bf B}_w^{-1} {\bf w}
\end{equation}
where ${\bf B}_w$ is the background error covariance matrix in the reduced space. The natural representation of ${\bf B}_w$ in the full space is the singular matrix 
\begin{equation}
{\bf B}_r = {\bf L} {\bf B}_w {\bf L}^T
\end{equation}
Minimization is performed using a quasi-Newton descent method with an
exact line search (algorithm M1QN3, Gilbert and Lemar\'echal 1989). As
in the classical 4D-Var method, the problem is preconditionned by defining a new
control variable $\delta {\bf v} = \textbf{B}^{-1/2} \delta{\bf x}_0$, which implies
$J_b(\delta {\bf v}) = \frac{1}{2}\, \delta {\bf v}^T \delta {\bf v}$.
From a programming point of view, this approach implies nearly no modification to the original code, since we only have to add a mapping procedure corresponding to ${\cal M}$, and the
adjoint of this procedure.  \par
It is important to point out that the choice of the subspace 
${\cal M}({\cal W})$ of ${\cal U}$ is performed using additional
information (the information leading to the construction of the ${\bf L}_i$s) with regard to usual 4D-Var with no order reduction. This is done of course in order to make the choice of ${\cal M}$ effective, but it will also automatically introduce this extra information into the assimilation procedure (through ${\bf L}$ and ${\bf B}_w$), and thus possibly help making the
assimilation efficient.\vspace*{3mm}\par
Concerning the actual choice of $({\bf L}_1,\ldots, {\bf L}_r)$, different families of vectors can be proposed :
\begin{itemize}
\item  The variability of the system can be defined in a statistical sense,
which means that we seek directions maximizing the variance around a mean
state of the system. This is actually the definition of Empirical Orthogonal
Functions (EOFs), which can be computed from a sampling of a model trajectory (see section \ref{SSect:eofs}).
\item We can also define the variability in a harmonical sense. In that case,
the vectors can be defined by a Fourier or wavelets analysis of a model
trajectory. Note however that, with regard to a rectangular domain, the presence of continental boundaries makes the analysis more difficult. 
\item If we consider the notion of variability within the framework of
dynamical systems, we look for vectors maximizing a ratio of the form
$\|{\bf x}(t=T_2)\| / \|{\bf x}(t=T_1)\|$, for some norm $\|.\|$. The problem can be simplified by making a
tangent linear approximation, which leads to the computation of singular
vectors (SVs). In the limit case where $T_2 - T_1$ becomes large (infinite),
SVs converge towards Lyapunov vectors (LVs). Properties of SVs and LVs can be
found for instance in Legras and Vautard (1995). The tangent linear
assumption can also be relaxed, and vectors corresponding to SVs and LVs can
be computed with the fully non-linear model. They are called respectively
non-linear singular vectors (NSVs, Mu 2000) and bred modes (BVs,
Toth and Kalnay 1997). Note that, to our knowledge, these ``non-linear'' vectors have been introduced in an empirical way, with nearly no related properties established theoretically.
\end{itemize}
Durbiano (2001) performed a thorough study of these families of
vectors (EOFs, SVs, LVs, NSVs and BVs) in the perspective of their use
as reduced basis for several data assimilation problem. In particular,
she compared their performances for the present problem of the control
of the initial condition in a reduced space, in the case of a 2-D
shallow water model. She concluded in this case to the clear
superiority of EOFs with regard to the other families of vectors. This
is probably due to the fact that EOFs take into account the
nonlinearity of the model (while SVs and LVs do not), and also that
their covariance matrix ${\bf B}_w$ is quite accurately known, which is not the case for the other families of vectors. 
That is why we used EOFs in the realistic 3-D experiment described in
section 3. Note that this way of approximating the variability of the system
in a data assimilation process by a low dimension space generated by the
first $r$ EOFs is similar to the method used in the SEEK filter, or in the reduced order filter proposed by Cane {\it et al.} (1996).  
\section{Numerical experiments}
\label{Sect:NumExp}
\subsection{Model and EOF analysis}
\label{SSect:eofs}
The model used in our tests is the primitive equation ocean general
circulation  model OPA (Madec {\it et al.} 1999), in its $z$-coordinate rigid-lid
version. The region of interest is the equatorial Pacific ocean, from 30\dg S
to 30\dg N. The horizontal resolution is set to 1\dg\/ zonally, and varies
meridionally from 1/2\dg\/ at the equator to 2\dg\/ at 30\dg. Vertically the
ocean is discretized using 25 levels. The state vector consists of
temperature, salinity and horizontal velocity,
and has a size slightly greater than $10^6$.\par  
A one-year simulation was performed, starting from
a previous restart built with the ECMWF wind stresses and heat fluxes
and using ERS-TAO daily wind stresses and ECMWF heat fluxes to force
the model. In a 10\dg-wide band near the northern and southern
boundaries, buffer zones are prescribed where the model solution is
relaxed towards Levitus climatology. This version of the model has
been used previously in a number of studies, and details can be found
therein (e.g. Vialard {\it et al.} 2001, Vialard {\it et al.} 2003,
Weaver {\it et al.} 2003).\par
The model solution during the first year of data assimilation experiment (1993) has been sampled with a 2-day periodicity, and a multivariate EOF analysis of the three-dimensional fields has been performed. Let us recall that this analysis consists in determining the main directions of variability of the model sample ${\bf X}=({\bf X}_1,\ldots,{\bf X}_p)$, which 
leads to diagonalizing  the covariance matrix ${\bf X}^T{\bf X}$,
with ${\bf X}_j= \displaystyle \frac{1}{\sigma_i} [{\bf x}(t_j)-\bar{\bf x}]$ and $\bar{\bf
x}=\displaystyle \frac{1}{p}\, \displaystyle \sum_{j=1}^{p}{\bf x}(t_j)$. The inner product is the usual one for a state vector containing several physical quantities expressed in different units~: 
\begin{equation}
<{\bf X}_j,{\bf X}_k> = \sum_{i=1}^{n} \displaystyle \frac{1}{\sigma_i ^2} ({\bf x}(t_j)-\bar{\bf x})_i ({\bf x}(t_k)-\bar{\bf x})_i
\end{equation}
where  ${\sigma_i^2}$ is the empirical variance of the $i$-th
component : ${\sigma_i^2}= \displaystyle \frac{1}{p}\,\displaystyle \sum_{j=1}^{p}({\bf X}_j^i)^2$.
This diagonalization leads to a set of orthonormal eigenvectors $({\bf
L}_1,\ldots, {\bf L}_p)$ corresponding to eigenvalues $\lambda _1 >
\ldots > \lambda_p >0$. Since trajectories are computed with
the fully non-linear model, these modes represent non-linear variability around
the mean state over the whole period.\par
The first level ($z=5$m) of the first EOF is displayed on Fig. 1. As can be
seen, it is mostly representative of the variability of the equatorial zonal
currents, of the north-south temperature oscillation and
of the mean structure of the sea surface salinity.\par
The fraction of variability (or ``inertia") which is conserved when
retaining only the $r$ first vectors is $\displaystyle \sum_{j=1}^{r}
\lambda_j / \sum_{j=1}^{p} \lambda_j$. Its variation as a function of
$r$ is displayed in Fig. 2. We can see that a large part of the total
variance can be represented by a very few EOFs : 80\% for the first 13 EOFs, 92\% for the first 30 EOFs.\par 
Finally, let us emphasize that a natural estimate for the covariance matrix of the first $r$ eigenvectors $({\bf L}_1,\ldots, {\bf L}_r)$, i.e. ${\bf B}_w$ in our reduced-order 4D-Var, is simply the diagonal matrix $\hbox{Diag}(\lambda_1,\ldots,\lambda_r)$.
\subsection{Assimilation experiments}
A 4D-Var assimilation scheme, based on the incremental formulation of
Courtier {\it et al.} (1994), has been developped for the OPA model
(Weaver {\it et al.} 2003, Vialard {\it et al.} 2003). Without going into details (which can be found in references above), let us recall that the nonquadratic cost function $J({\bf x}(t_0))$ is expressed in terms of the increment $\delta {\bf x}_0$, and that its minimization is replaced by a sequence of minimizations of simplified quadratic cost functions. The basic state-trajectory used in the tangent linear model is regularly updated in an outer loop of the assimilation algorithm, while the iterations of the actual minimizations are performed within an inner loop.\par
Different statistical models can be chosen for representing the
correlations of background error. In the present study, we used a
Laplacian-based correlation model, which is implemented by numerical
integration of a generalized  diffusion-type equation (Weaver and
Courtier, 2001). The horizontal correlation lengths for the gaussian
functions are equal to $8^o$ in longitude and $2^o$ in latitude near the
equator and $4^o$ in longitude/latitude outside the area situated between $20^o$N/S. The vertical
correlation lengths depend on the depth. ${\bf B}$ is thus block diagonal :
covariances are spatially varying but remain monovariate.  Such a choice for ${\bf B}$ leads to significantly
better results than those given by a simple diagonal representation of
this matrix. However, since ${\bf
B}$ remains univariate, the links between the model variables
come only from the action of the model dynamics. The development of a
multivariate model for ${\bf B}$ is presently under way in research
groups. Ricci {\it et al.} (2004) include a state-dependent
temperature-salinity constraint, which works quite well in the 3D-Var
case but is not yet operational for the 4D-Var case.\par
The observation error covariance matrices ${\bf R}_i$ depend of course of the assimilated data. 
We will consider in the present case only temperature observations,
which are assumed independent with a standard error equal to $\sigma_T$.
The ${\bf R}_i$ are thus taken equal to $\sigma_T^2\, {\bf Id}$.\par   

We have used for our experiments the classical framework of
twin experiments. A one-year simulation of the model was performed, starting
at the beginning of 1993.  This
simulation (further denoted $E_{REF}$) will be the reference experiment. 
Pseudo-observations of the temperature field were then generated, by extraction
from this one-year solution at the locations of the 70 TAO moorings (Fig.
3), with a periodicity of 6 hours, on the first 19 levels of the model (i.e.
the first 500 meters of the ocean). This corresponds to observing 0.17\% of the model state vector every 6 hours. Those temperature values have been perturbed by the addition of a gaussian noise, with a standard error set to $\sigma_T=0.5$\dg C, which is an upper bound for the standard error of
the real TAO temperature dataset.\par     
A 4D-Var assimilation of these pseudo-observations (i.e. with full
control variable $\delta {\bf x}_0$, built from the state vector
(u,v,T,S) in the whole space) was then performed,
using an independent field ${\bf x}_b$ (a solution of the model three
months later) as the first guess (background field) for the minimization
process. This first assimilation experiment will be denoted $E_{FULL}$, since it uses the full control space. In order to improve the validity of the tangent linear approximation, the assimilation time window was divided into successive one-month windows.\par    
Then an additional simulation was performed, using the reduced-space
approach described in section 2 with $r=30$ EOFs (which represent 92\% of the
total inertia - Fig. 2). This second assimilation experiment will be denoted $E_{REDUC}$. As detailed previously, the control variable in this case
is ${\bf w}=(w_1, \ldots, w_r)$, with the mapping $\delta {\bf x}_0 =
{\bf L}{\bf w}$ and the preconditionning $\delta {\bf v} = {\bf B}_w^{-1/2}{\bf w} = {\bf B}_w^{-1/2}{\bf L}^T \delta{\bf x}_0$.
\subsection{Numerical results}
As explained in section 2, the reduced-space assimilation algorithm presents two main differences with regard to the full-space algorithm, which are the   multivariate nature of the background error covariance matrix, and the small dimension of the control space. Both aspects are expected to improve the efficiency of the assimilation, and we will now illustrate their respective impact.
\subsubsection{Background error covariances}
The background error covariance matrix used in the reduced-space approach is defined empirically by the EOF analysis and is expressed in the full-space as  ${\bf B}_r={\bf L}{\bf B}_w{\bf L}^T$. It integrates statistical information on the consistency between the different model variables, and is naturally multivariate. On the other hand, the matrix ${\bf B}$ used in the full-space 4D-Var is univariate, since providing a multivariate model for this matrix remains challenging. This aspect is of course very important, and should lead to significant changes in the assimilation results. Note that Buehner {\it et al.} (1999) have proposed a similar way of representing error covariances with EOF analysis in the context of 3D-Var. However they consider that the reduced basis is not sufficient to span the analysis increment space and blend this EOF basis  with the prior ${\bf B}$ projected into the sub-space orthogonal to the EOFs.\par
An interesting way to illustrate these differences between the
full-space ${\bf B}$ and the reduced-space ${\bf B}_r$ is to perform
preliminary assimilation experiments with a single observation. For
that purpose, we use a single temperature observation located within
the thermocline at 160\dg W on the equator, and specified at the end
of a one-month assimilation time window. The innovation is set to 1\dg
K. The analysis increment at the initial time in such an experiment is
proportional to the column of ${\bf B}{\bf M}_{t_n,t_0}^T$
corresponding to the location of the observation. As can be seen in
Fig. 4, the reduced-space method performs, as expected,  a rather weak correction over the whole basin, while the full-space method generates a much stronger and local increment.
The structure of the increment is indeed much more elaborate in the
reduced-space experiment, with scales larger than in the full-space experiment. Note that the input from the first EOF (shown on Fig. 1) is quite clear in the horizontal pattern of the increment, since $w_1/\|{\bf w}\|=0.86$ in this particular case. The maximum value of the increment however is only 0.06\dg C for the reduced-space 4D-Var, while it is 0.94 \dg C in the full-space 4D-Var.\par
The interest of the naturally multivariate aspect of  ${\bf B}_r$ is also clear in the results of our twin experiments. Two different types of diagnostics were performed, the first one concerning only the assimilated variables (i.e. temperature in the present case), while the second one relates to all other variables that are not assimilated. This second type of diagnostic is of course the most significant, since it evaluates the capability of the assimilation procedure to propagate information over the whole model state vector.\par
An example of the first type of diagnostic is given in Fig. 5a, which displays the temperature rms error defined by 
\begin{equation}
\hbox{rms}_T (z,t)=\left(\int{
\left( T(\lambda,\theta,z,t)-T_{REF}(\lambda,\theta,z,t)\right)^2  d\lambda \,
d\theta } \right)^{1/2}
\end{equation}

The discretized formula becomes :

\begin{equation}
\hbox{rms}_T (z,t)= \| x -x_{ref} \|_2 = \left [\displaystyle \frac{1}{N_x \times N_y}
\sum_{i=1}^{N_x} \sum_{j=1}^{N_y} (\textbf{T}(i,j,z,t) -
\textbf{T}_{ref}(i,j,z,t))^2\right ]^{1/2} 
\end{equation}
where $N_x$ and $N_y$ are the number of grid points in x and y.
This error is significantly weaker in $E_{REDUC}$ than in $E_{FULL}$, although the assimilation system in $E_{REDUC}$ has much less degrees of freedom to adjust the model trajectory to these data.\par 
An example of the second type of diagnostic is shown in Fig. 5b,c.  In our test case, these results are clearly in favour of
the reduced-space approach. The errors on the salinity S and the zonal component of the velocity $u$ for the solution
provided by $E_{FULL}$ are systematically greater than for $E_{REDUC}$. \par
The interest of this approach can also be illustrated by the results in the
lower levels. It is well-known that the time-scale for the information to penetrate from the upper ocean into the deep ocean within an
assimilation process may be quite long. However, in experiment $E_{REDUC}$ the
EOFs add information on the vertical structure of the flow (see Fig. 4) and then make the
vertical adjustment easier. We have plotted for
example in Fig. 6 the errors of the different solutions at level 20 (depth :
750 m, \textit{ie} below the observations). $E_{REDUC}$ performs a
very good identification of the solution due to the propagation of the information in depth.\par
These results are only part of what should be shown in terms of
diagnostic analyses. But all of them clearly prove that the results of
$E_{REDUC}$ vs $E_{FULL}$ are significantly improved for all, assimilated or not, variables.\par 
Finally, it must be mentioned that we have also illustrated the fundamental role of the multivariate nature of ${\bf B}_r$ by performing an additional reduced-order experiment (not shown) using univariate EOFs. In this case, the directions proposed for the minimization were not relevant, and the assimilation failed. 
\subsubsection{Dimension of the control space}
The second important difference brought by the reduced-space approach with regard to the full-space approach is the dimension of the minimization space, which is decreased by several orders of magnitude. This should reduce the number of iterations necessary for the minimization, i.e. reduce the cost of the data assimilation algorithm, which is an important practical issue.\par 
The evolution of the cost functions for experiments $E_{FULL}$ and
$E_{REDUC}$ are displayed on Fig. 7. Since we use different covariance matrices ${\bf B}$ and ${\bf B}_r$ in these two experiments, the curves are not quantitatively
comparable.
However, it is clear in Fig. 7 that the number of iterations required to
stabilize the cost function is reduced by nearly one order of
magnitude between the full-space 4D-Var approach (which needs
typically several tens of iterations) and the reduced-space approach
(which needs eight to ten iterations). In the present experiments, we
have kept the same number of iterations (2 outer loops of ten
iterations each) in the two experiments to strictly compare the
results. But having a look at the cost function, it is clear that the
minimum is quickly reached by $E_{REDUC}$ experiment. Considering the
low number of freedom degrees, the computational cost can be thus divided by a factor of 4 or 5 between the two methods.
\section{Conclusion}
This paper presents a reduced-space approach for 4D-Var data assimilation. A new control space of low dimension is defined, in which the minimization is performed. An illustration of the method is given in the case of twin experiments with a primitive equation model of the equatorial Pacific ocean. \par
This method presents two important features, which make the assimilation algorithm effective. First the background error covariance matrix ${\bf B}_r$ is built using statistical information (an EOF analysis) on a previous model run. This introduces relevant additional information in the assimilation process and makes ${\bf B}_r$ naturally multivariate, while providing an analytical multivariate model for ${\bf B}$ is still challenging. This improves the identification of the solution, both on observed and non-observed variables, and at all depths in the model.
Secondly the reduction of the dimension of the control space limits the number of iterations for the minimization, which results in a decrease of the computational cost by roughly one order of magnitude.
\par
However the results presented in this
work are only a first (but necessary) step, since they concern twin experiments. They need of course to be confirmed by
additional experiments in other contexts, in particular experiments with real data and in other geographical areas. 
As a matter of fact, the efficiency of the method is closely related
to the fact that the reduced basis does contain pertinent information on the
variability of the true system. That is why, in the context of real
observations (i.e. in the case of an imperfect model), the control
space must probably not be limited to model-based variability.
Therefore, we can imagine either compute EOFs from results of previous
data assimilation using for example full-space 4D-Var (Durbiano 2001),
and/or improve the assimilation results by performing a few full-space iterations at the end of the reduced-space minimization (Hoteit {\it et al.} 2003).\par
Several other ideas can be considered to extend the present methodology to a fully realistic context, and some of them are presently under investigation in our group. Concerning the definition of the reduced basis, one could think of its evolutivity and adaptivity, as in some sequential assimilation methods (Brasseur {\it et al.} 1999; Hoang {\it et al.} 2001). Moreover a major source of difficulty (common to all data assimilation methods) is our insufficient knowledge (and therefore parameterization) of the model error. Recent works have addressed this problem in the context of variational methods, which intend to model and control this error (e.g. D'Andr\'ea and Vautard 2001; Durbiano 2001; Vidard 2001). Such a control could probably be performed in a reduced-order context and complement efficiently the present method. 
\vspace*{5mm}\\
{\bf Acknowledgments}\par
The authors would like to thank Anthony Weaver and Arthur Vidard for numerous helpful discussions. A. Weaver provided the OPAVAR package and helped us using it. Laurent Parent helped in the configuration of the numerical experiments. This work has been supported by the french project MERCATOR for operational oceanography. Idopt is joint
CNRS-INPG-INRIA-UJF research project.\vspace*{3 mm}\\ 
%
%

%
%
\newpage
%
%
\begin{figure}
\begin{center}
{\includegraphics[height=0.45\linewidth,width=0.8\linewidth]{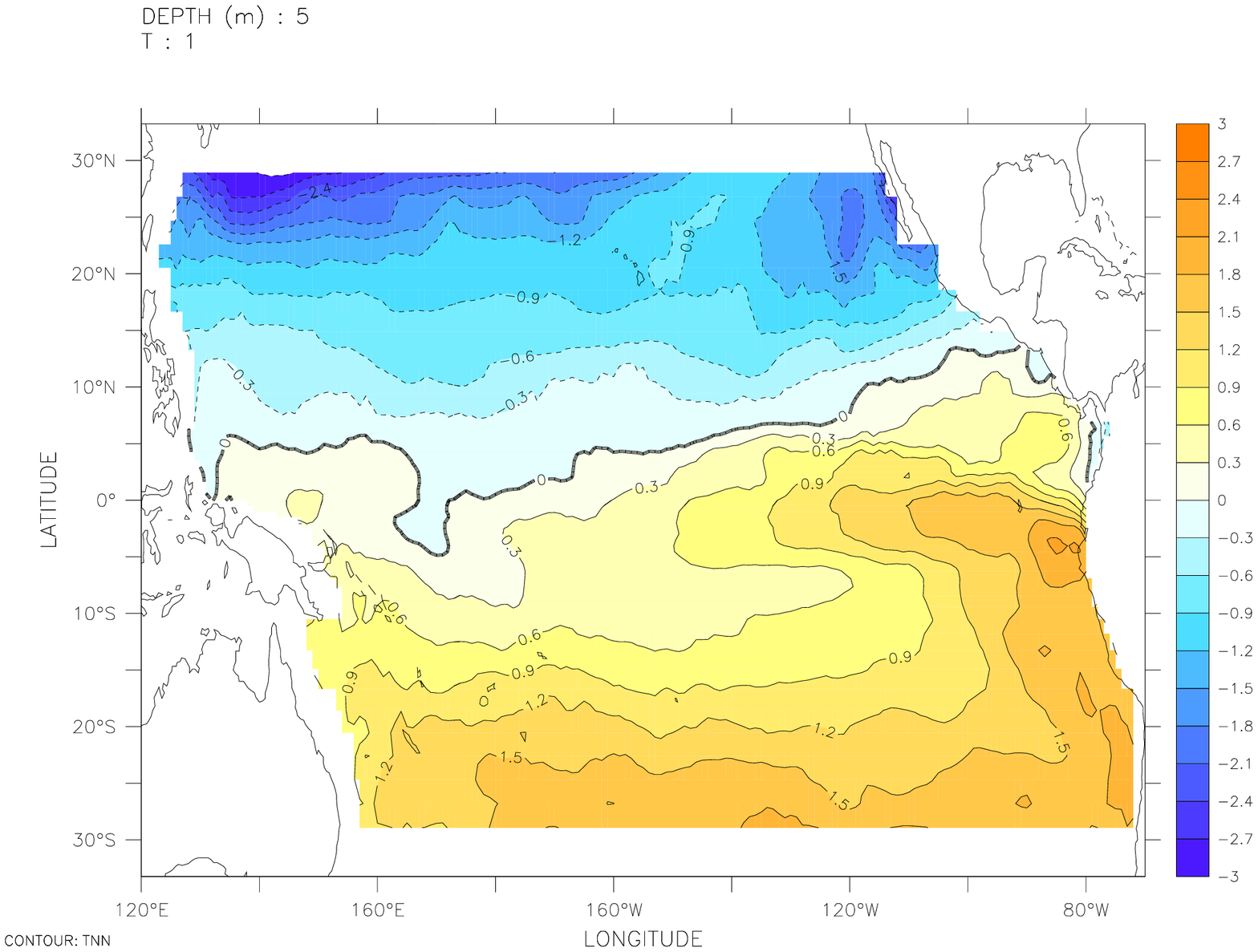}}\\
{\includegraphics[height=0.45\linewidth,width=0.8\linewidth]{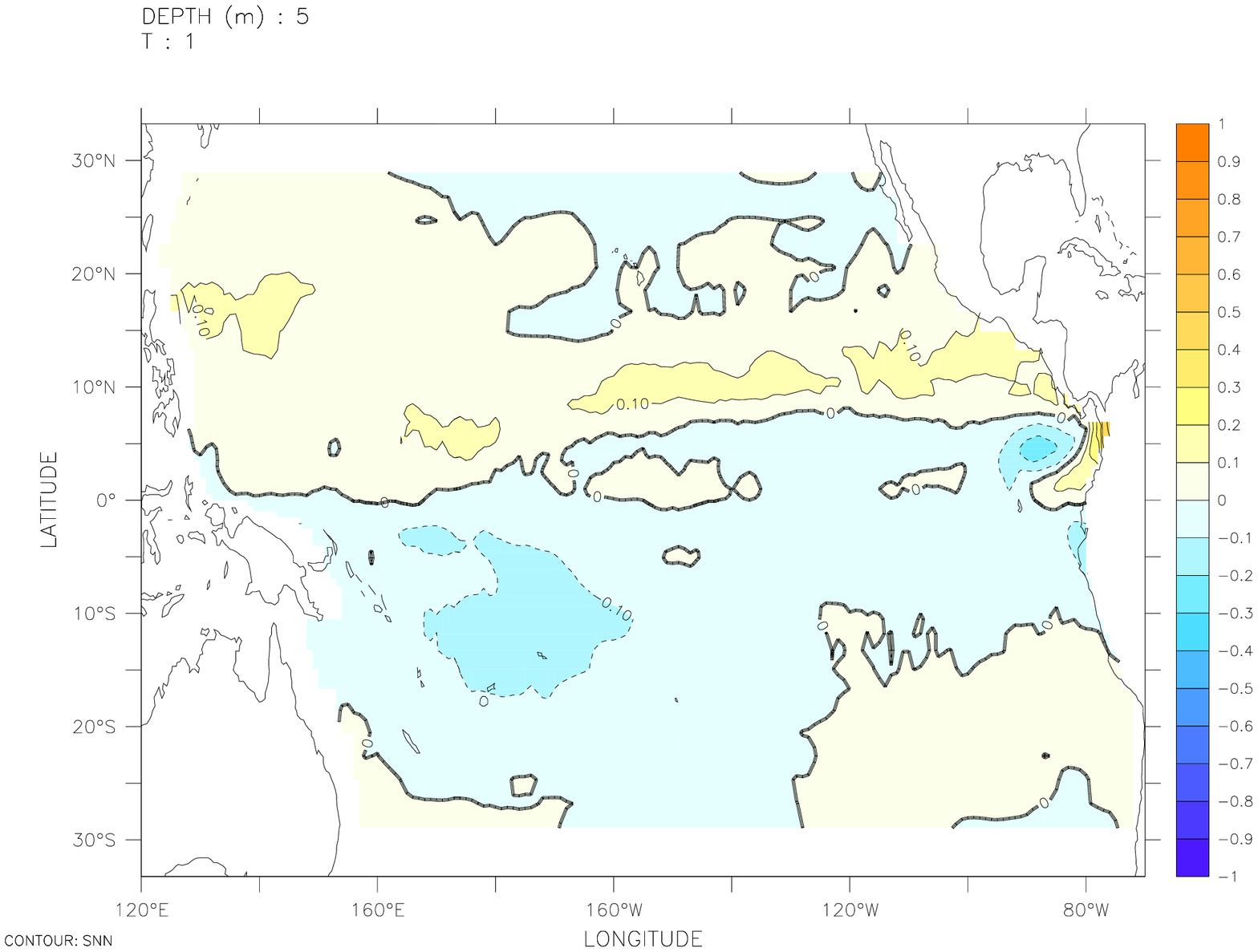}}\\
{\includegraphics[height=0.45\linewidth,width=0.8\linewidth]{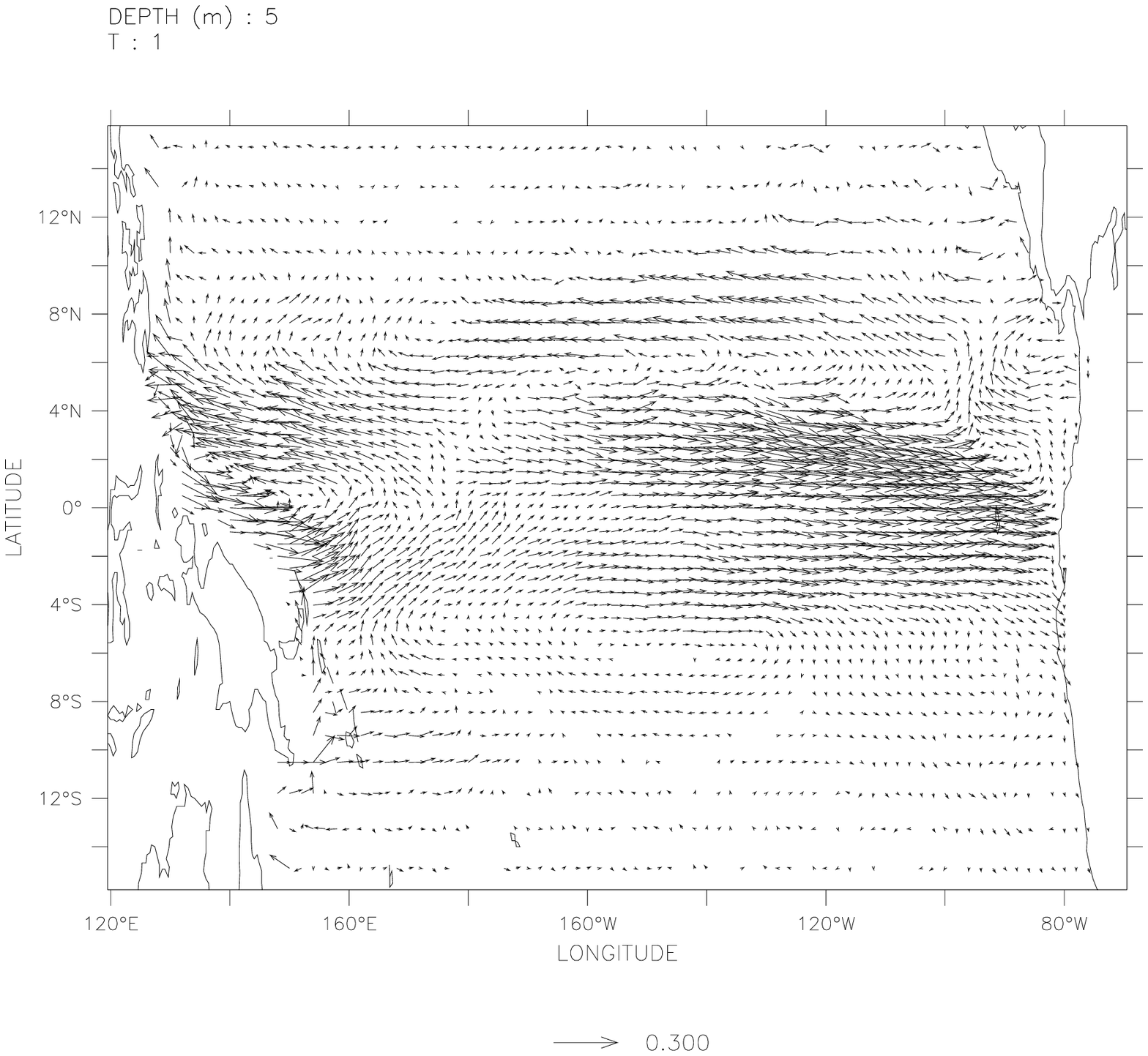}}
\end{center}
\caption{First EOF. Top: surface temperature; Middle: surface salinity; Bottom: surface
velocity. The quantities are non-dimensional.} 
\end{figure}
\clearpage
%
%
\begin{figure}
\rotatebox{-90}{\includegraphics[width=0.7\linewidth]{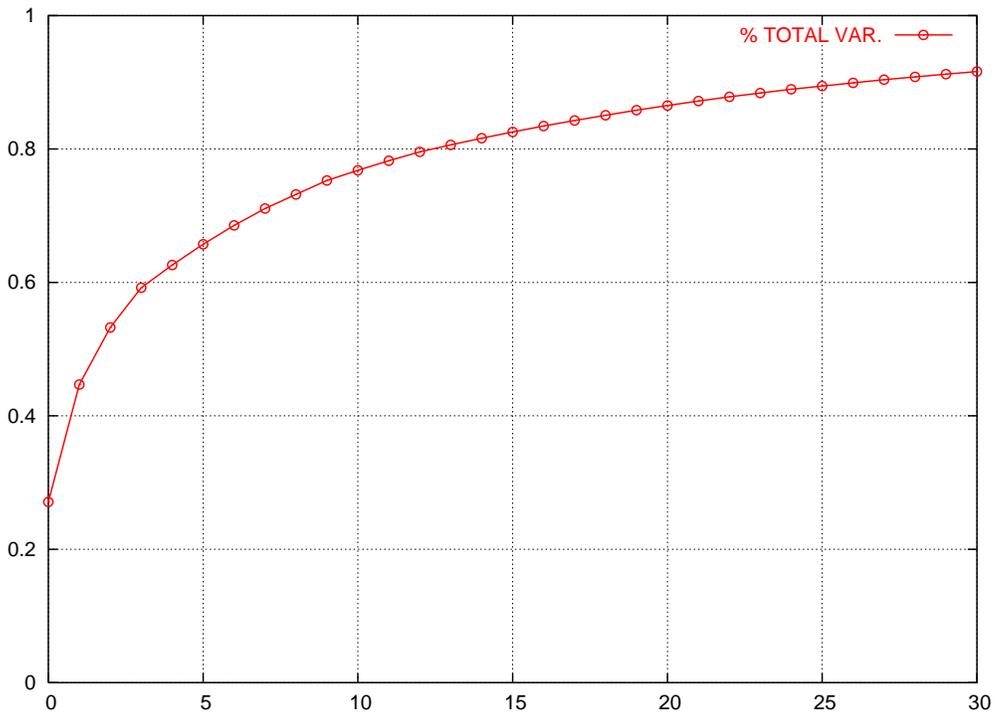}}
\caption{Fraction of inertia conserved by the $r$ first EOFs : $\sum_{j=1}^{r} \lambda_j / \sum_{j=1}^{p} \lambda_j$ as a function of $r$}
\end{figure}
\clearpage
%
%
\begin{figure}
\includegraphics[width=0.45\linewidth,angle=-90]{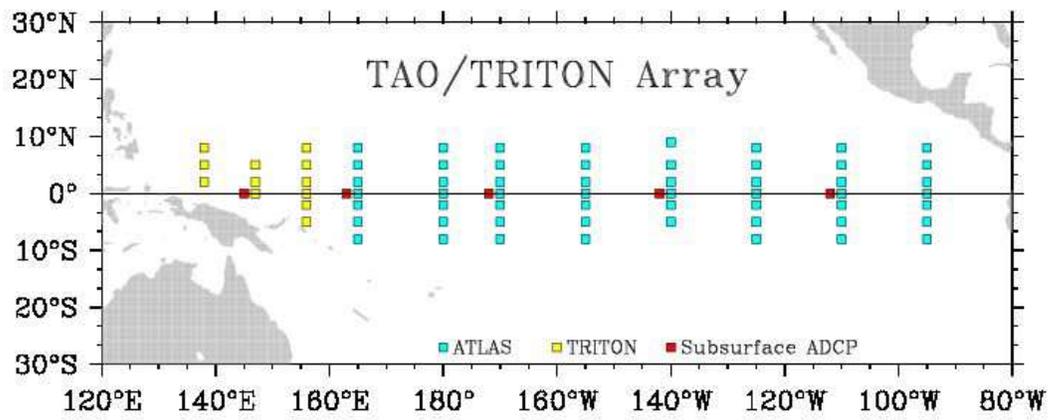}
\caption{Locations of the TAO morrings.}  
\end{figure}
\clearpage
%
%
%
\begin{figure}
\begin{tabular}{ll}
\includegraphics[width=0.48\linewidth]{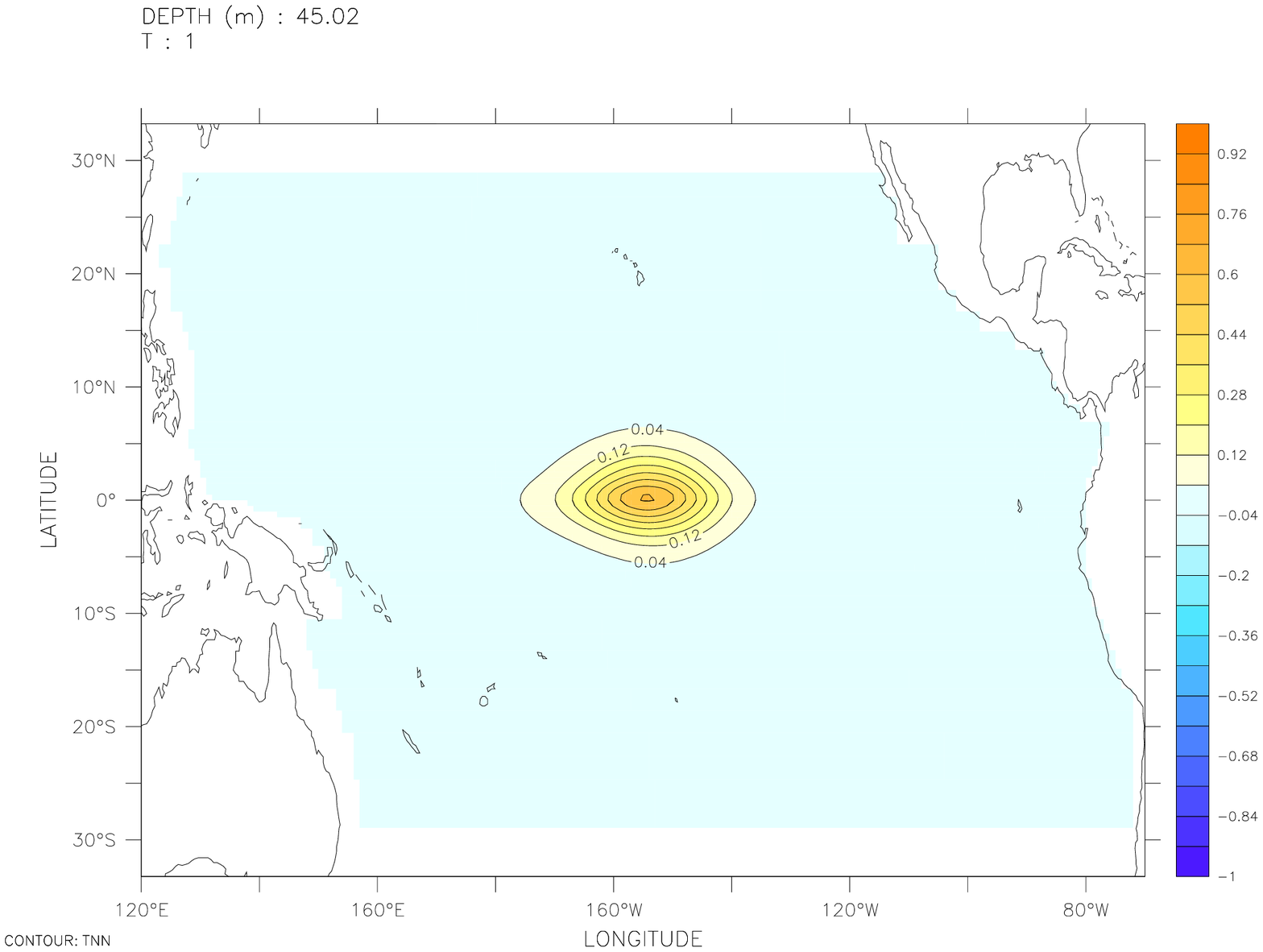} &
\includegraphics[width=0.48\linewidth]{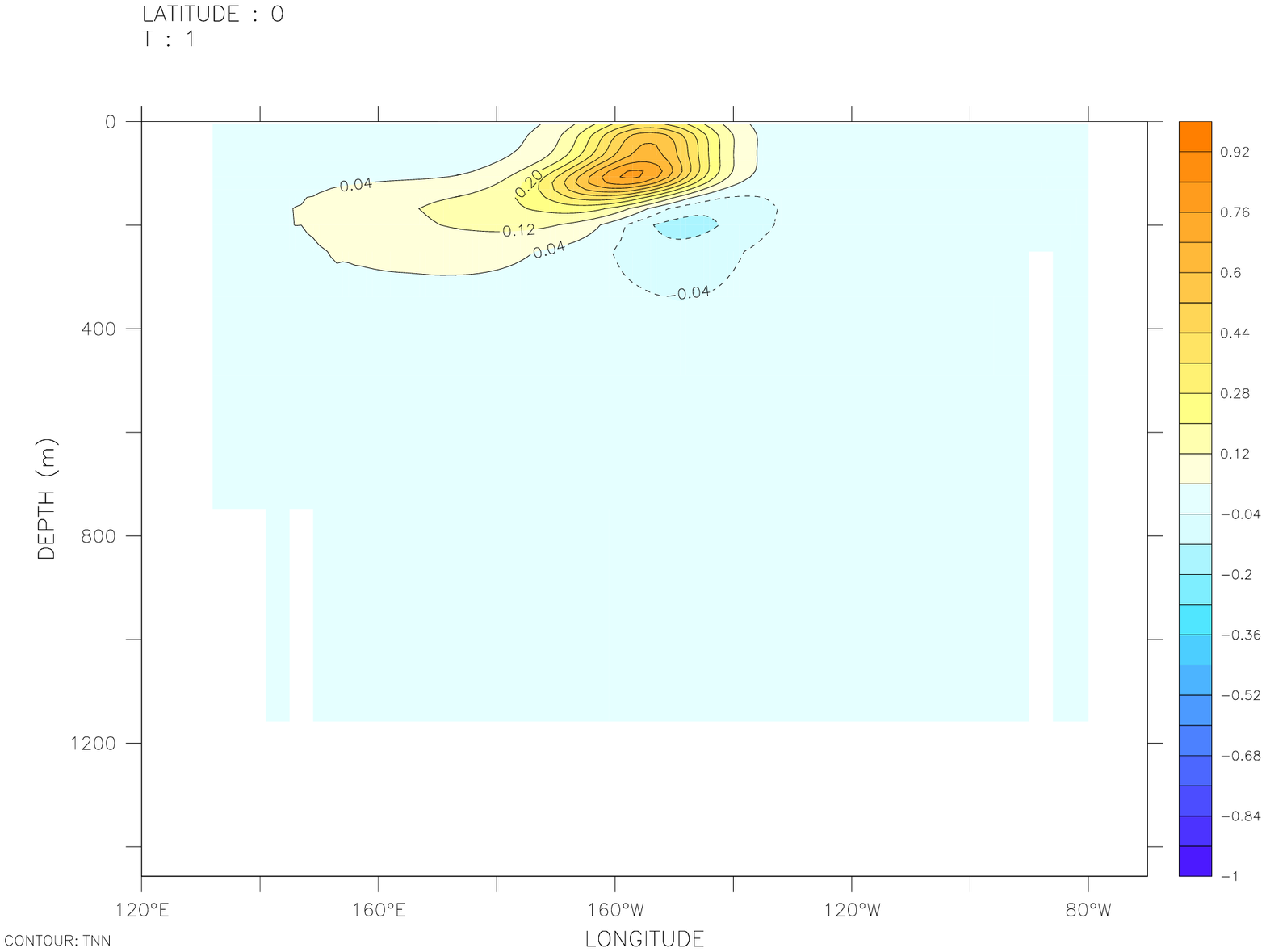} \\
\includegraphics[width=0.48\linewidth]{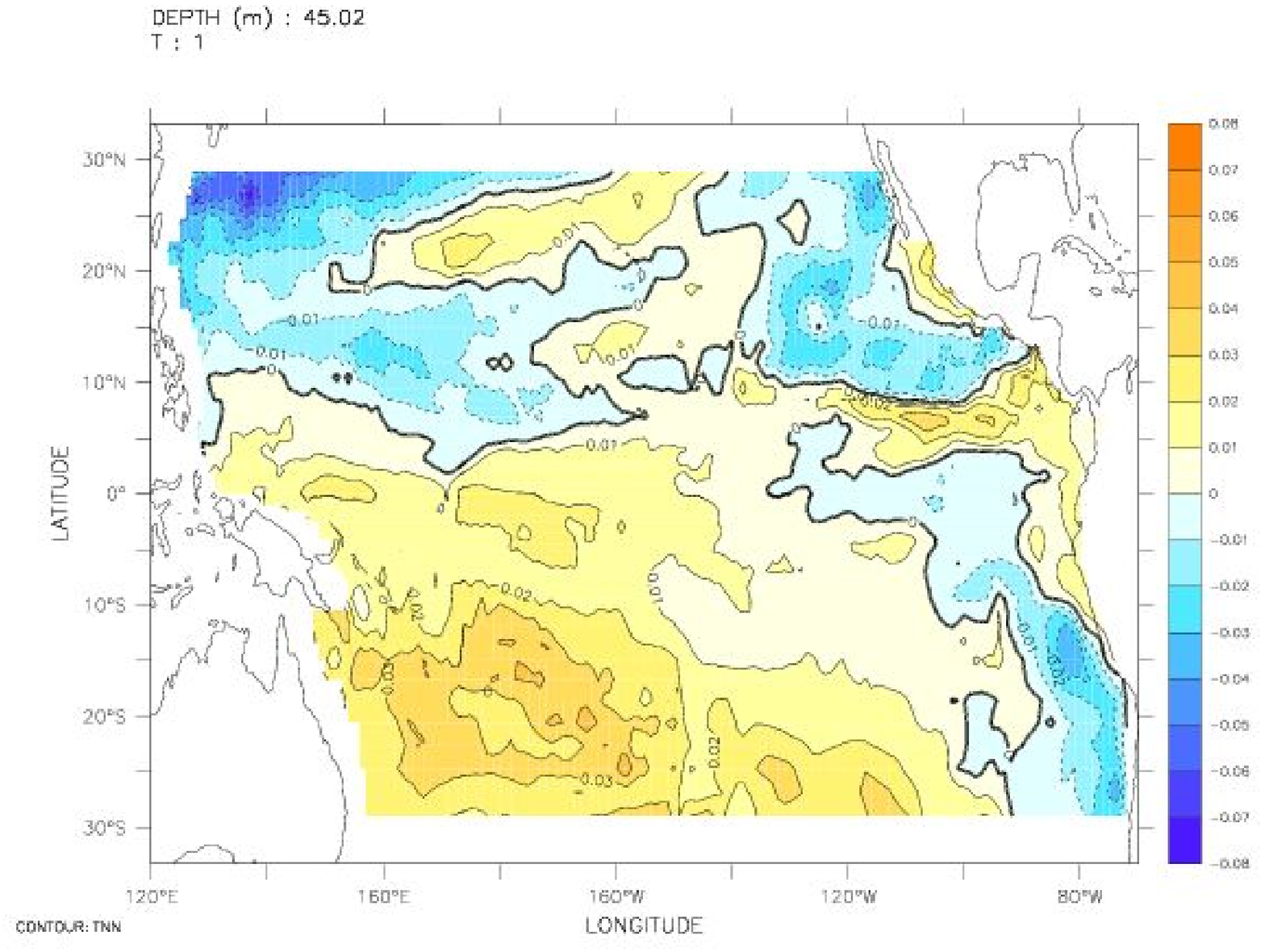} &
\includegraphics[width=0.48\linewidth]{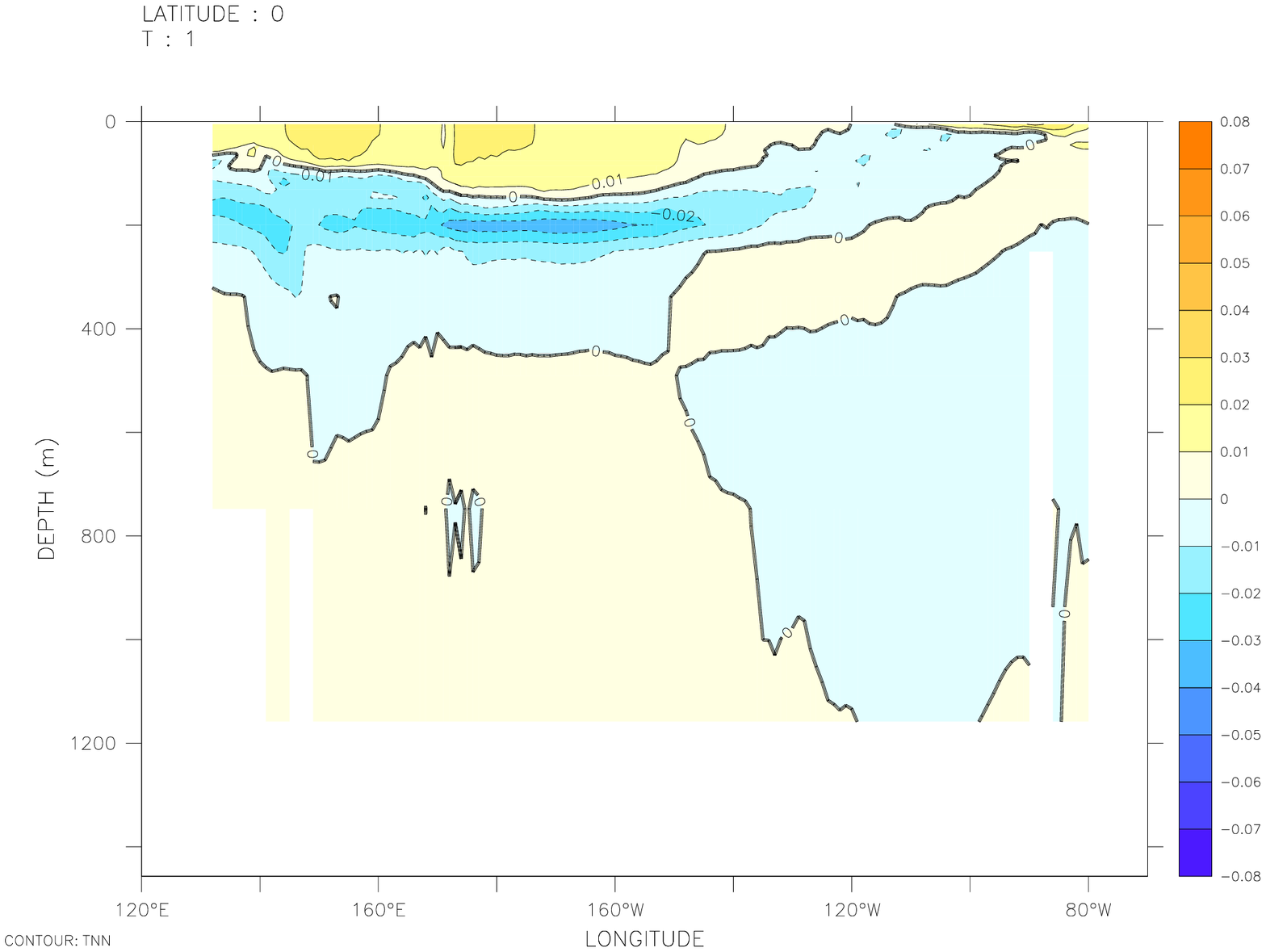} \\
\end{tabular}
\caption{Temperature component of the optimal increment $\delta {\bf x}_0$ for single observation experiments. Left : horizontal structure at $z=-45$ m; right : vertical section along the equator. Top : full-space 4D-Var; bottom : reduced-space 4D-Var.}  
\end{figure}
\clearpage
%
%
\vspace{-0.5cm}
\begin{figure}
\vspace{-0.5cm}
\begin{tabular}{c}
 \includegraphics[width=0.45\linewidth,angle=-90]{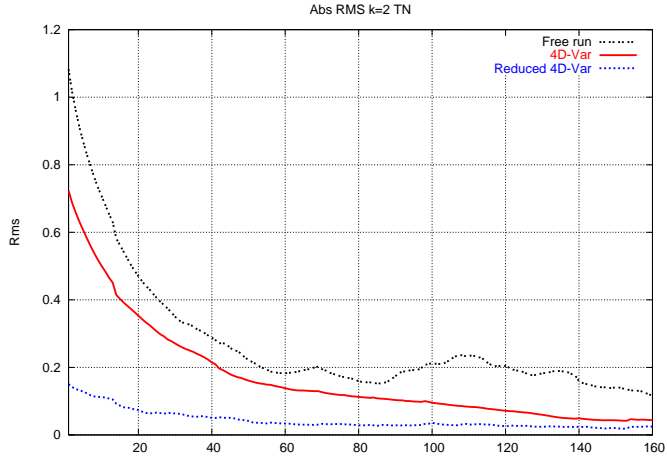} \\
a) \\
 \includegraphics[width=0.45\linewidth,angle=-90]{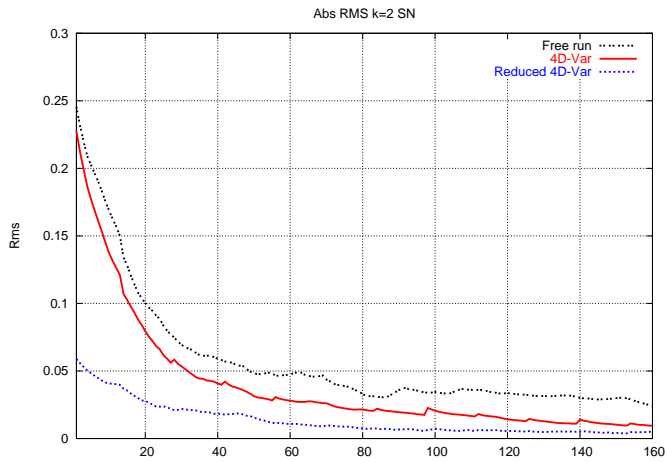} \\
 b) \\
 \includegraphics[width=0.45\linewidth,angle=-90]{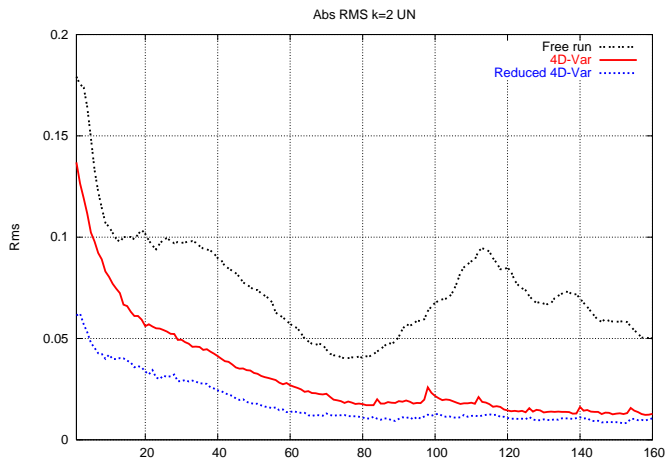} \\
 c) \\
\end{tabular}
\caption{Rms error with respect to the exact reference
solution at level 2 (depth: 15 m). $x$-axis : time (in days). $y$-axis : (a) $T$ ($^\circ$K), (b) $S$
(kg.m$^{-3})$, (c) $u$ (m.s$^{-1}$). The curves correspond to experiment $E_{REF}$ (dotted line), $E_{FULL}$ (solid line) and $E_{REDUC}$ (dotted line).}  
\end{figure} 
\clearpage
%
%
\begin{figure}
\begin{tabular}{c}
\includegraphics[width=0.45\linewidth,angle=-90]{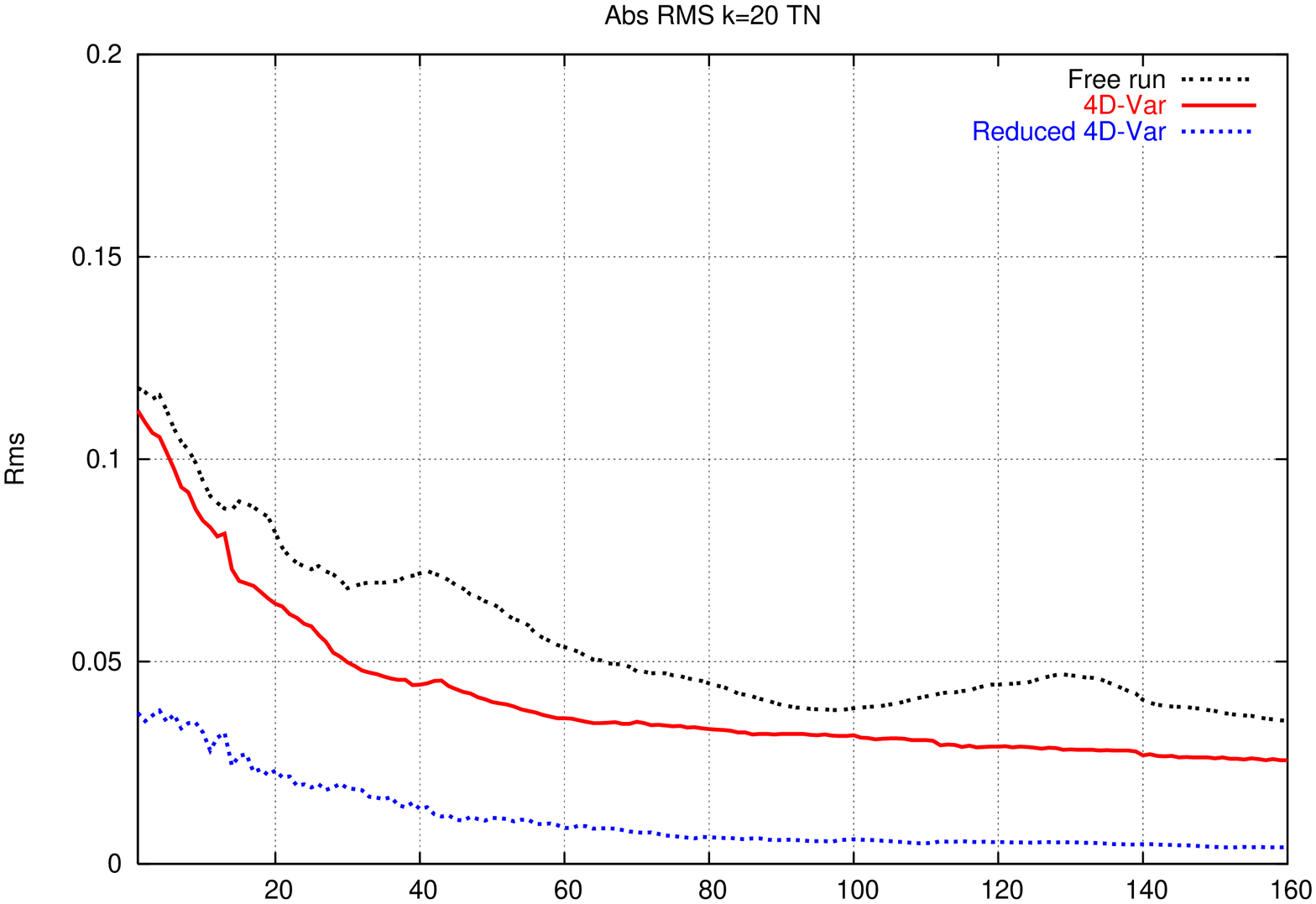} \\
a) \\
\includegraphics[width=0.45\linewidth,angle=-90]{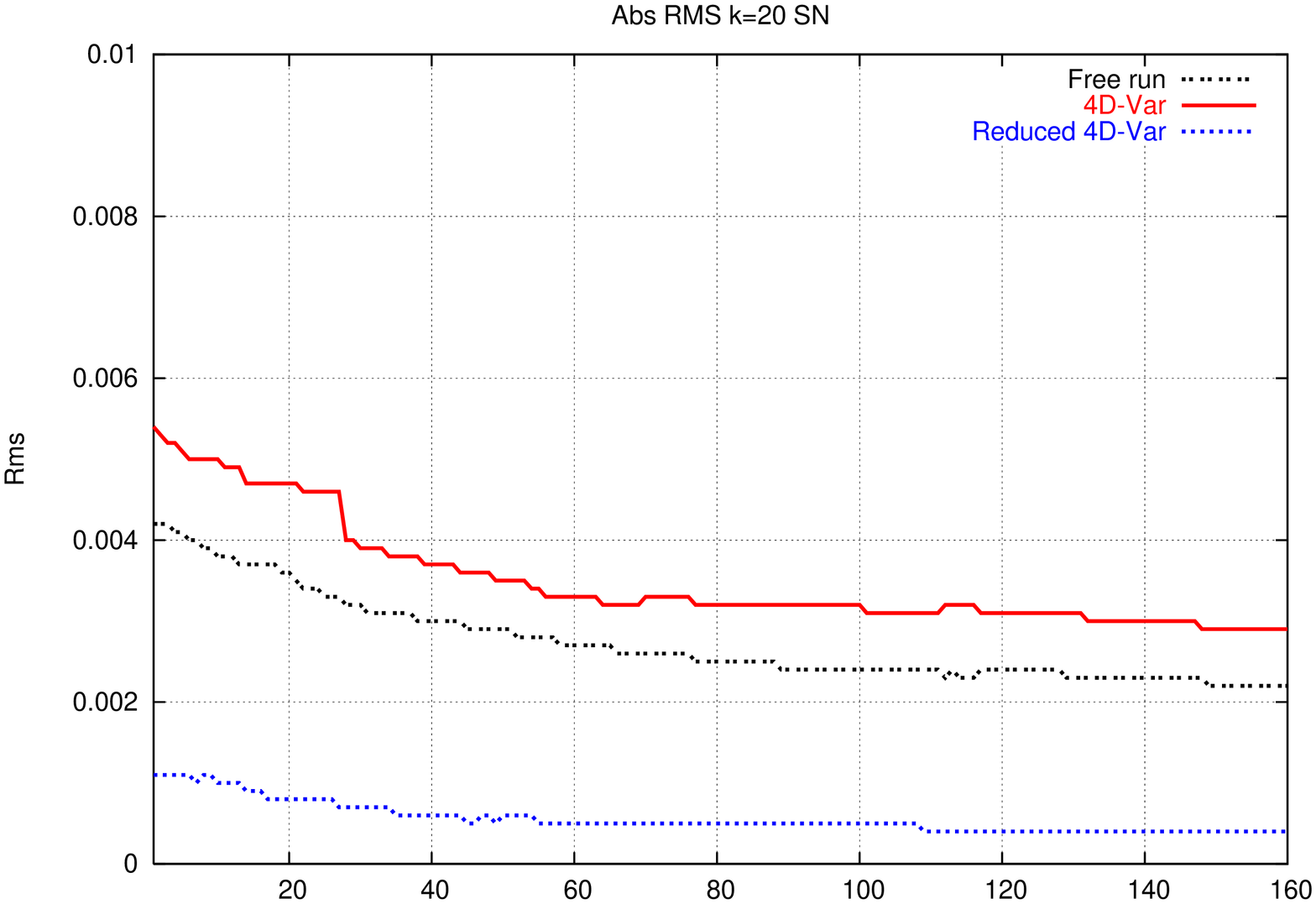} \\
b) \\
\includegraphics[width=0.45\linewidth,angle=-90]{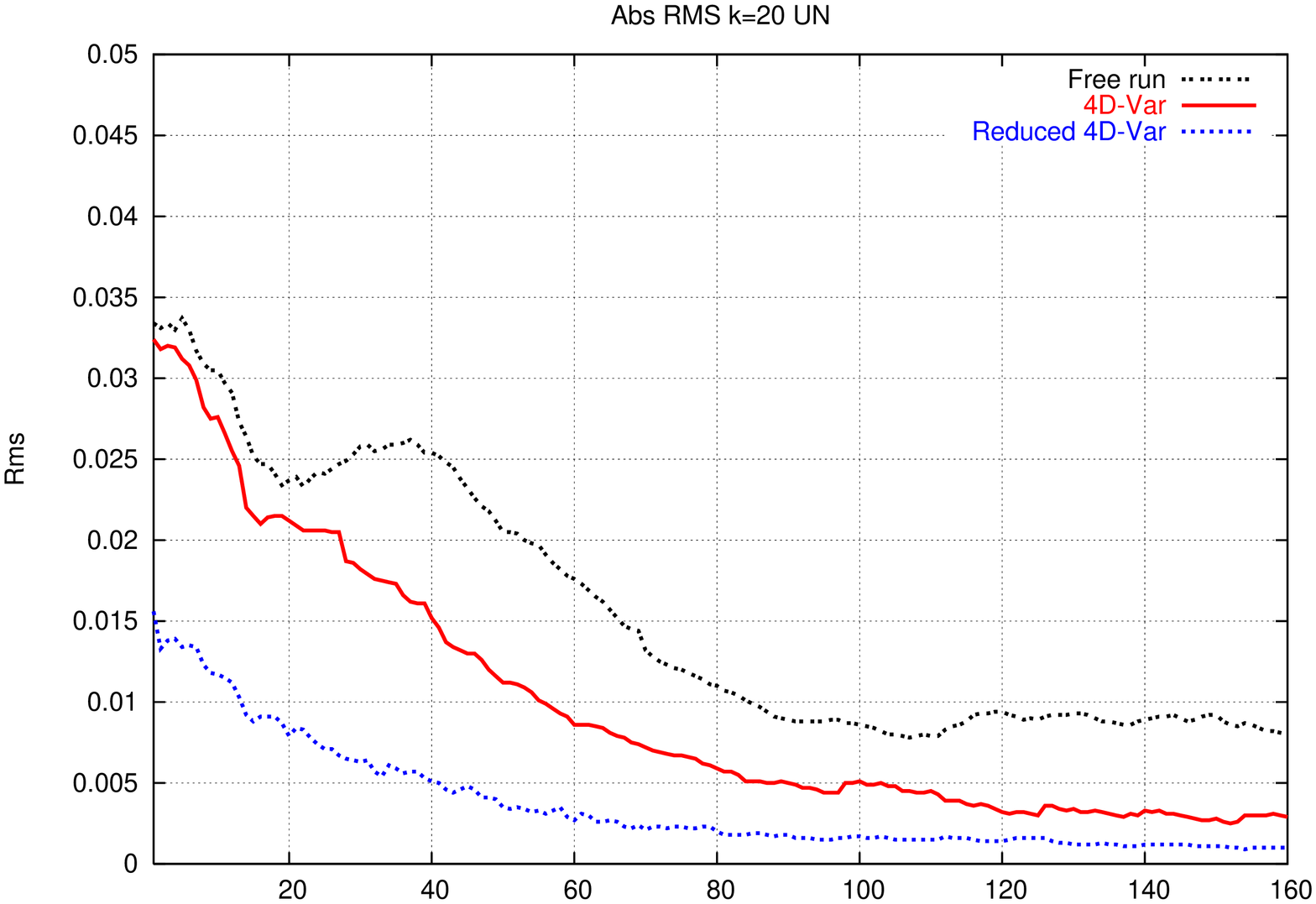} \\
c) \\
\end{tabular}
\caption{Same as Fig. 5, but at level 20 (depth: 750 m).} 
\end{figure}
\clearpage
%
%
\begin{figure}
\begin{center}
\rotatebox{-90}{\includegraphics[width=0.75\linewidth,angle=90]{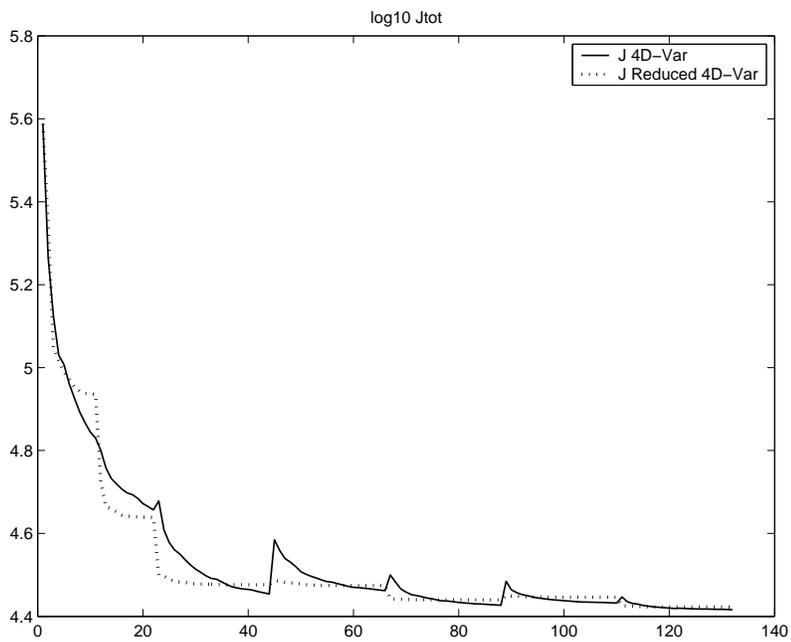}}
\end{center}
\caption{Cost functions vs iterations. Solid line: experiment $E_{FULL}$ (22 iterations for each of the six one-month assimilation time-windows); Dotted line: experiment $E_{REDUC}$ (22 iterations for each of the six one-month assimilation time-windows)}  
\end{figure}
\end{document}